\documentclass[copyright,creativecommons]{eptcs}
\providecommand{\volume}{??}
\providecommand{\anno}{??}
\usepackage{breakurl}

\usepackage{amsmath,amssymb}
\usepackage{amsthm}
\usepackage{subfigure}
\usepackage{latexsym}
\usepackage{graphicx} 
\usepackage[english]{babel} 
\usepackage{semantic}
\usepackage{mycommands}
\usepackage{preamble}
\usepackage{xspace} 
\usepackage{color} 
\usepackage{longtable} 
\usepackage{proof}
\usepackage{myproof}

\def\midd{\; \; \mbox{\Large{$\mid$}}\;\;}

\newcommand{\Guarded}{\cG} 
\newcommand{\hide}[1]{}

\title{Towards a Unified Framework for Declarative Structured Communications}
 \author{Hugo A. L\'opez
\institute{IT University of Copenhagen}
\email{lopez@itu.dk}\\
\and
Carlos Olarte
\institute{\'Ecole Polytechnique \quad Universidad Javeriana Cali}
\email{colarte@lix.polytechnique.fr}
 \and
Jorge A. P\'erez
\institute{University of Bologna}
\email{perez@cs.unibo.it}\\
}

\def\titlerunning{Towards a Unified Framework for Declarative Structured Communications}
\def\authorrunning{L\'opez, Olarte, \& P\'erez}

\begin{document}
\maketitle

\begin{abstract}
We present a unified framework for
the declarative analysis of structured communications.
By relying on a (timed) concurrent constraint programming language, we
show that in addition to the usual operational techniques from process
calculi, the analysis of structured communications can elegantly exploit logic-based reasoning
techniques. 
We introduce
a declarative 
interpretation of the language for structured communications proposed
by Honda, Vasconcelos, and Kubo.
Distinguishing features of
our approach are: 
 the possibility of including partial information (constraints) in the session model; 
the use of explicit time for reasoning about session duration and expiration; 
a tight correspondence with logic, which formally relates session
execution and linear-time temporal logic  formulas. 
\end{abstract}

\section{Introduction}
\paragraph{Motivation.}
From the viewpoint of \emph{reasoning techniques}, two main trends 
in modeling in Service Oriented Computing (SOC) can be singled out.
On the one hand, an \emph{operational approach} focuses on how process interactions can lead to correct configurations.
Typical 
representatives of this approach 
are based on process calculi and Petri nets (see, e.g., \cite{vanderaalst1998apn,boreale2006ssc,lanese2007doa,lapadula7cows}), and count with behavioral equivalences and type disciplines as main analytic tools. 
On the other hand, in a \emph{declarative approach} 
the focus is on the set of conditions  components should fulfill in order to be considered correct, 
rather than on the complete specification of the control flows within process activities (see, e.g., \cite{vanderaalst2006dtt,pesic2006daf}). 
Even if these two trends address similar concerns, we find that they have evolved rather independently from each other.

The quest for a unified approach in which 
operational and declarative techniques can harmoniously converge is therefore a legitimate research direction. 
In this paper we shall argue that Concurrent Constraint Programming (CCP) \cite{cp-book} can serve as a foundation for such an approach. 
Indeed, the unified framework for operational and logic techniques 
that CCP provides can be 
fruitfully exploited for analysis in SOC, possibly in conjunction with other techniques such as type systems.
Below we briefly introduce the \ccp model and then elaborate 
on how it can shed light on a particular issue: the analysis of 
structured communications.

\ccp\  \cite{cp-book} is a well-established model for concurrency where processes  interact with each other by \emph{telling} and \emph{asking} for pieces of information (\emph{constraints}) in a shared medium, the \emph{store}. 
While the former operation simply adds a given constraint to the store (thus making it available for other processes),
the latter allows for rich, parameterizable forms of process synchronization. 
Interaction is thus inherently \emph{asynchronous}, and can be related to a broadcast-like communication discipline, as opposed to the point-to-point discipline enforced by formalisms such as the $\pi$-calculus \cite{SaWabook}. 
In \ccp, the information in the store grows monotonically, as constraints  cannot be removed.  
This condition is relaxed in  \emph{timed} extensions of \ccp (e.g., \cite{tcc-lics94,NPV02}), where  processes evolve along a series of \emph{discrete time units}. 
Although each unit contains its own store, information is not automatically transferred from one unit to another. 
In this paper we shall adopt a \ccp process language that is timed in this sense.

In addition to the traditional operational view of process calculi, \ccp enjoys a 
\emph{declarative} nature that distinguishes  it from other models of concurrency:   \ccp programs can be  seen, at the same time, as computing agents and as logic formulas \cite{cp-book,NPV02,Olarte:08:PPDP}, i.e.,  they can be read and understood as logical specifications. 
Hence, 
CCP-based languages are 
suitable for \emph{both} the specification and 
verification 
of programs. In the \ccp\ language used in this paper, processes can be interpreted as linear-time temporal logic formulas;  
we shall exploit this correspondence to verify properties of our models.

\paragraph{This Work.} 
We describe initial results on the definition of a formal framework for the declarative analysis of structured communications.
We shall exploit \utcc \cite{Olarte:08:SAC}, a  timed CCP process calculus, to give 
a declarative 
interpretation to the language  defined by Honda, Vasconcelos, and Kubo in  \cite{honda1998lpa} (henceforth referred to as \hvk).
This way, structured communications can be analyzed in a declarative framework where 
time  is defined explicitly.
 We begin by proposing an encoding of the \hvk language  into \utcc 
and studying its correctness.
 We then move to the timed setting, and propose \hvkplus, a timed extension of \hvk. 
 The extended language explicitly includes information on session duration,
 allows for declarative preconditions within session establishment constructs,
 and features a construct for session abortion.
 We then discuss how the encoding of \hvk into \utcc straightforwardly extends to \hvkplus.

\paragraph{A Compelling Example.}
We now give intuitions on how a declarative approach could be useful in the analysis of structured communications.
Consider the ATM example from \cite[Sect. 4.1]{honda1998lpa}. 
There, an ATM has established two sessions:
the first one with a user, sharing session $k$ over service $a$, 
and the second one with the bank, sharing session $h$ over service $b$.
The ATM offers \texttt{deposit}, \texttt{balance}, and \texttt{withdraw} operations.
When executing a withdraw, if there is not enough money in the account, then an \emph{overdraft} message appears to the user. 
It is interesting to analyze what occurs when 
this scenario is extended 
to consider a 
card reader that acts as a malicious interface between the user and the ATM. The user communicates his personal data with the reader using the service $r$, which will be kept by the reader after the first withdraw operation  
to continue withdrawing money without the authorization of the user.
A greedy card reader could even withdraw repeatedly  until causing an overdraft, as expressed below:

{\footnotesize \centering
\begin{tabular}{lcl} \\ 
	${\it Reader}$	&  =	& $ \accept{r}{k'} \receive{k'}{id}$\\
			&	&	\quad $ \request{a}{k} \send{k}{id} $
					\qquad $\branching{k'}{ \begin{array}{l}
											withdraw: \receive{k'}{amt}  \\
											~~\selection{k}{withdraw} \send{k}{amt} \\
												      \quad \branching{k}{dispense : \selection{k'}{dispense} \send{k}{amt} R(k,amt) || overdraft: Q}
											\end{array}
											}$\\
	${\it R(j,x)}$	& = &	$\recursionH{R'} \selection{k}{withdraw} \send{j}{x} \branching{j}{dispense : \receive{j}{amt} R' || overdraft: Q}$\\
	${\it User}$	&  =	& $ \request{r}{k'} \send{k'}{myId}$\\
			&	&	\quad $\selection{k'}{withdraw} \send{k'}{58}$
			\tab $\branching{k'}{dispense : \receive{k'}{amt} P || overdraft: Q}$ \\ \\ 
\end{tabular}
}

By creating sessions between them, the card reader $Reader$ is able to receive the 
user's information, and to use 
it later by attempting a session establishment with the bank. 
Following authentication steps (not modeled above),  
the card reader allows the user to obtain the requested amount. 
Additional withdrawing transactions between the reader and the bank are defined by the recursive process $R$.
In the specification above, the process $Q$ can be assumed to send a message (through a session with the bank)
representing the fact that 
the account has run out of money: $Q = \send{k_{bank}}{\underline{\mathsf{0}}} \inact$. 

Even in this simple scenario, 
the combination of operational and declarative reasoning techniques 
may come in handy  to reason about 
the possible states of the system.
Indeed, while an operational approach
can be used to describe an operational description of the compromised ATM above, 
the declarative approach can complement such a description by 
offering declarative insights regarding its evolution. 
For instance, assuming $Q$ as above, 
one could show that a \utcc specification of the ATM example 
satisfies  the linear temporal logic formula $\sometime \outp(k_{bank}, 0)$, 
which intuitively means that in  presence of a malicious card reader 
the user's bank account will eventually reach an overdraft status.

\paragraph{Related Work.}
One approach to combine the declarative flavor of constraints and process calculi techniques is represented by a number of works that 
have extended name-passing calculi with some form of partial information (see, e.g., \cite{VictorP98,DiazRV98}). 
The crucial difference between such a strand of work 
and \ccp-based calculi
is that the latter offer a tight correspondence with logic, which greatly broadens the spectrum of reasoning techniques at one's disposal.
Recent works similar  to ours include CC-Pi \cite{BuscemiM07} and the calculus for structured communications in \cite{CD08}. 
Such languages feature elements that resemble much ideas underlying CCP (especially \cite{BuscemiM07}). 
The main difference between our approach and such works is that 
we adhere to the use of declarative reasoning techniques based on temporal logic 
as an effective way of complementing operational reasoning techniques. 
In \cite{BuscemiM07}, 
the reasoning techniques associated to CC-Pi are essentially operational, and used to reason about service-level agreement protocols. 
In \cite{CD08},  
the key for analysis is represented by a type system which provides consistency for session execution, much as in the original approach in \cite{honda1998lpa}.

\section{Preliminaries}
\label{s:preli}
\subsection{A Language for Structured Communication}
We begin by introducing \hvk, a language for structured communication proposed in  \cite{honda1998lpa}. We assume the following conventions: 
\emph{names} are ranged over by $a, b, \dots$; 
\emph{channels} are ranged over by $k, k'$; 
\emph{variables} are ranged over by $x, y, \dots$; 
\emph{constants} (names, integers,  booleans) are ranged over by $c, c' ,\dots$; 
\emph{expressions} (including constants) are ranged over by $e, e', \dots$; 
\emph{labels} are ranged over by $l, l', \dots$; 
\emph{process variables} are ranged over by $X, Y , \dots$. 
Finally, $ u, u', \dots$ denote names and channels. We shall use $\vec{x}$
to denote a sequence (tuple) of variables $x_1...x_n$ of length $n=|\vec{x}|$. 
Notation $\vec{x}$ will be similarly applied to other syntactic entities.  The sets of free names/channels/variables/process variables of $P$, is defined in the standard way, and are respectively denoted by $\fn(\cdot)$, $\fc(\cdot)$, $\fv(\cdot)$, and $\fpv(\cdot)$.  Processes without free variables or free channels are called \emph{programs}.

\begin{definition}[The \hvk language \cite{honda1998lpa}] \label{def-hondasLanguage}
Processes in \hvk are built from:

 {\small
  \begin{longtable}{rrllrrll}
	P,Q	&$::=$	& 	\request{a}{k}P 	&	Session Request 
		&$|$	&	\accept{a}{k}P	&	Session Acceptance\\
		&$|$	&	\send{k}{\vec e}P	&	Data Sending 
		&$|$	&	\receive{k}{\vec x}P	&	Data Reception\\
		&$|$	&	\selection{k}{l}P	&	Label Selection
		&$|$	&	\branching{k}{l_1:P_1 \parallel \cdots || l_n:P_n}     &   Label Branching      \\
		&$|$	&	\throw{k}{k'}P	&	Channel Sending 
		&$|$	&	\catch{k}{k'}P	&	Channel Reception \\
		  &$|$ &		\itn{e}{P}{Q}	& 	  Conditional Statement 
		  &$|$ &		P $|$ Q		&	  Parallel Composition\\
		  &$|$ &		\inact		&	  Inaction 
		  &$|$ &		$(\nu u)$P		&	  Hiding\\
		  &$|$ &		\recursionH{D}P	& Recursion
		  &$|$ &		$X[\vec{e} \, \vec{k} ]$	& Process Variables\\
	D	  &$::=$ &		\multicolumn{2}{l}{$X_1(x_1 k_1) = P_1$ $\mathbf{and} \cdots \mathbf{and}$ $X_n(x_n k_n) = P_n$}\\
			& &\multicolumn{2}{r}{Declaration for Recursion}  
 \end{longtable}
}
 \end{definition}

\paragraph{Operational Semantics of \hvk.}
The operational semantics of \hvk\ is given by the reduction relation $\redihvk$ which is the smallest relation on processes generated by the rules in Figure \ref{tab:sos-hvk}.   In Rule $\textsc{Str}$, the structural congruence $\equiv _h$ is the smallest relation satisfying : 
1) $P \equiv _h Q$ if they differ only by a renaming of bound variables (alpha-conversion). 2) $P \ |\  \inact \equiv _h P$, $P \ |\  Q \equiv _h Q \ |\  P$, $(P \ |\  Q) \ |\  R \equiv _h P \ |\  (Q \ |\  R)$. 3)  $(\nu u) \inact \equiv _h \inact$,  $(\nu u u')P \equiv _h (\nu u' u)P$, $(\nu u )(P \ |\  Q) \equiv _h (\nu u)P \ |\  Q$ if $x \notin \fv(Q)$, $(\nu u)(\recursionH{D}P) \equiv _h (\recursionH{D}((\nu u)P)) $ if $u \notin \fv(D)$. 4) $(\recursionH{D}P) \ |\ Q \equiv _h \recursionH{D}(P \ |\  Q)  $ if $\fpv(D)\cap \fpv(Q) = \emptyset$. 5) $\recursionH{D}{(\recursionH{D'}P)} \equiv _h \recursionH{D \mbox{ and } D'}P$ if $\fpv(D)\cap \fpv(D') = \emptyset$.

\begin{figure}[h]
{\small
\[\begin{array}{ll}
{\textsc{Link}} & \request{a}{k}Q \ |\  \accept{a}{k}P  \redihvk (\nu k)(P \ |\  Q ) \\
\textsc{Com} &  (\send{k}{\vec e}P) \ |\  (\receive{k}{\vec{x}}Q ) \redihvk P \ |\  Q [\vec{c}/\vec{x}] \mbox{\quad if } e\downarrow \vec{c}\\
\textsc{Label} & \selection{k}{l_i}P \ |\  \branching{k}{l_1:P_1 \parallel \cdots \ \parallel   l_n:P_n} \redihvk  P \ |\  Pi \ (1 \leq i \leq n) \\
\textsc{Pass}  &  \throw{k}{k'}P	\ |\  \catch{k}{k'}Q \redihvk P \ |\  Q \\
\textsc{If1}   & \itn{e}{P}{Q} \redihvk P \ (e \downarrow  \true)  \\
\textsc{If2}   & \itn{e}{P}{Q} \redihvk Q\ (e \downarrow  \false)  \\
\textsc{Def} &  \recursionH{D}{(X[\vec{e} \, \vec{k} ] \ |\  Q )} \redihvk  
\recursionH{D}{(P[\vec{c}/\vec{x}] \ |\  Q) } \  (e \downarrow \vec{c}, X(\vec{x}\vec{k}) = P \in D)\\
\textsc{Scop} & P \redihvk P' \mbox{ implies } (\nu u)P \redihvk (\nu u)P' \\ 
\textsc{Par} & P \redihvk P'  \mbox{ implies }  P \ |\  Q \redihvk  P' \ |\  Q  \\
\textsc{Str} & \mbox{If } P \equiv _h P'  \mbox{ and } P' \redihvk Q'  \mbox{ and } Q' \equiv _h Q \mbox{ then } P \redihvk Q
\end{array}\]
}
\caption{Reduction Relation for \hvk\  ($\redihvk$)\cite{honda1998lpa}. \label{tab:sos-hvk}}
\end{figure}

Let us give some intuitions about the language constructs and the rules in Figure \ref{tab:sos-hvk}. 
The central idea in \hvk\ is the notion of a \emph{session}, i.e., a series of reciprocal interactions between two parties, possibly with 
branching, delegation and recursion, which serves as an abstraction unit for describing structured communication. 
Each session has associated a specific port, or \emph{channel}. Channels are generated at session initialization; communications inside the session take place on the same channel. 

More precisely, sessions are initialized by 
a process of the form 
$ \request{a}{k}Q \ |\  \accept{a}{k}P$. 
In this case, there is a request, on name $a$, for the initiation of a session and the 
generation of a fresh channel. 
This request is matched by an accepting process on $a$, which 
generates a new channel $k$, thus allowing $P$ and $Q$ to communicate each other. 
This is the intuition behind rule $\textsc{Link}$.
Three kinds of atomic interactions are available in the language: 
sending (including name passing), branching, and channel passing (also referred to as delegation). 
Those actions are described by rules $\textsc{Com}$, $\textsc{Label}$, and $\textsc{Pass}$, respectively. 
In the case of $\textsc{Com}$,
the expression $\vec{e}$ is sent on the port (session channel) $k$. 
Process $\receive{k}{\vec{x}}Q $ then receives such a data and executes $Q[\vec{c}/\vec{x}]$, 
where  $\vec{c}$ is the result of evaluating the expression $\vec{e}$. 
The case of  $\textsc{Pass}$ is similar but  considering that in the constructs $\throw{k}{k'}P$ and $\catch{k}{k'}Q$, only session names can be  transmitted. 
In the case of 
$\textsc{Label}$, the process $\selection{k}{l_i}P$ selects one label and then the corresponding process $P_i$ is executed. 
The other rules are self-explanatory. 


For the sake of simplicity, and without loss of generality (due to rule 5 of $\equiv _h$),  in the sequel we shall assume programs of the form $\recursionH{D}P$  where there are not  procedure definitions in $P$.

\subsection{Timed Concurrent Constraint Programming}\label{sec:utcc}
Timed concurrent constraint programming (\tcc) \cite{tcc-lics94} extends CCP 
for modeling reactive systems. In \tcc, time is conceptually divided into 
\emph{time units} (or \emph{time intervals}). In a particular time
unit, a \tcc process $P$  gets an input (i.e. a constraint) $c$
from the environment, it executes with this input as the initial
\emph{store}, and when it reaches
its resting point, it \emph{outputs} the resulting store $d$ to the
environment. The resting point determines also a residual process $Q$
which is then executed in the next time unit. 
It is  worth noticing that the final store is not automatically transferred to the next time unit. 

The \utcc  calculus  \cite{Olarte:08:SAC} extends \tcc for reactive systems featuring mobility.
Here \emph{mobility} is understood as 
the dynamic reconfiguration of system linkage through communication, 
much like in the $\pi$-calculus \cite{SaWabook}.    
\utcc generalizes \tcc by considering 
a \emph{parametric} ask operator of the form $\absp{\vec{x}}{c}{P}$, with the following intuitive meaning:
process $P[\vec{t}/\vec{x}]$ is executed for every term $\vec{t}$ such that the current store entails  an admissible substitution $c[\vec{t}/\vec{x}]$. 
This process can be seen as an \emph{abstraction} of the process $P$ on the variables $\vec{x}$ under the constraint (or with the \emph{guard}) $c$.

\utcc\  provides a number of reasoning techniques:
First, \utcc processes can be represented as partial closure operators (i.e. idempotent and  extensive functions). Also, for 
a significant fragment of the calculus,
the input-output behavior of a process $P$  can be retrieved from the set of fixed points of its  associated closure operator \cite{Olarte:08:PPDP}. Second,  \utcc\ processes can be  characterized as  
First-order Linear-time Temporal Logic (FLTL)
formulas  \cite{mp91}. This declarative view of the processes  allows for the use of the  well-established verification techniques from FLTL to reason about \utcc\ processes. \\

\noindent{\bf Syntax}. 
Processes in \utcc\  are parametric in a \emph{constraint  system}  \cite{cp-book} which specifies 
the basic constraints 
that agents can tell or ask during execution.  
It also defines an \emph{entailment} relation ``$\entails$'' specifying interdependencies among constraints. 
Intuitively, $c \entails d$ means that the information in $d$ can be deduced from that in $c$ (as in, e.g., $x > 42 \entails x > 0$). 

The notion of constraint system can be set up by using first-order logic (see e.g., \cite{NPV02}).  We assume a first-order signature $\Sigma$ and a (possibly empty) first-order theory $\Delta$, i.e., a set of sentences over $\Sigma$ having at least one model. Constraints are then first-order formulas over 
$\Sigma$. Consequently, the entailment relation is defined as follows: $c\entails d$ if the implication $c\Rightarrow d$ is valid in $\Delta$. 

The syntax of the language is as follows:
  \[
{
\begin{array}{ll}
P,Q := & \skipp  \midd  \tellp{c}  \midd \absp{\vec{x}}{c}{P} \midd  P\parallel Q \midd \localp{\vec{x} ; c}{P}  \midd \nextp{P}  \midd  \unlessp{c}{P}  \midd  \bangp{P}
 \end{array}
 }
\]  with the variables in $\vec{x}$ being pairwise distinct. 

A process $\skipp$ does nothing; a process $\tellp{c}$ adds $c$ to the store in the current time interval. A process $Q=\absp{\vec{x}}{c}{P}$ binds the variables $\vec{x}$ in $P$ and $c$. It executes $P[\vec{t}/\vec{x}]$ for every term $\vec{t}$ s.t. the current store entails an admissible substitution over $c[\vec{t}/\vec{x}]$. The substitution $[\vec{t}/\vec{x}]$ is admissible if $|\vec{x}| = |\vec{t}|$ and no  $x_i$ in $\vec{x}$ occurs in $\vec{t}$. 
Furthermore, $Q$ evolves into $\skipp$ at the end of the time unit,  i.e.,  abstractions are not persistent when passing from one time unit to the next one.    
  $\parp{P}{Q}$ denotes $P$ and $Q$ running in parallel during the current time unit. A process \(\localp{\vec{x};c}{P} \)  \emph{binds}  the variables $\vec{x}$ in $P$ by declaring them  private to $P$ under a constraint $c$. If $c=\true$, we  write $\localp{\vec{x}}{P}$ instead of $\localp{\vec{x};\true}{P}$.
The \emph{unit delay} \( \nextp{P} \) executes $P$ in the next time unit.  The \emph{time-out} \( \unlessp{c}{P} \) is also a unit delay, but  \( P \)  is executed in the next time unit iff \( c \) is not entailed by the final store at the current time unit. Finally, the \emph{replication} \( !\,P \) means   \( P\parallel \mathbf{next}\,P\parallel \mathbf{next}^{2}P\parallel
\dots\), i.e., 
an unbounded number of
copies of \( P \) but one at a time. 
We shall use $\bangp_{[n]}P$ to denote  
\emph{bounded replication}, i.e., $P\parallel \nextp{P}\parallel ... \parallel \nextp^{n-1}P$. 

 From a programming language perspective, variables $\vec{x}$ in  $\absp{\vec{x}}{c}{P}$ can be seen as the formal parameters of $P$. This way, \emph{recursive definitions} of the form   $X(\vec{x}) \defsymbol P$ can be encoded in \utcc as
 \begin{equation}\label{eq:def-encoding}
	 \cR\os X(\vec{x}) \defsymbol P \cs =  \bangp\absp{\vec{x}}{{\it call_x}(\vec{x})}{\widehat{P}}
\end{equation}
 where ${\it call_x}$ is an uninterpreted predicate (a constraint) of arity $|\vec{x}|$.
 Process  $\widehat{P}$ is obtained from  $P$ by replacing  recursive calls of the form $X(\vec{t})$  with $\tellp{{\it call_x}(\vec{t})}$. Similarly,  calls of the form $X(\vec{t})$ in other processes are replaced with $\tellp{{\it call_x}(\vec{t})}$.    
     
\paragraph{Operational Semantics.}
The operational semantics  considers \emph{transitions}  between process-store \emph{configurations}  \(
\left\langle P,c\right\rangle  \)  with stores  represented as constraints and processes  quotiented by the structural congruence $\equiv_u$ defined below. We shall use $\gamma,\gamma',\ldots$ to range over configurations. 

The semantics is given in terms of an \emph{internal} and an \emph{observable} transition relation; both are given in Figure \ref{opersem}. 
The \emph{internal transition}  { \( \left\langle P,d\right\rangle \redi \left\langle P',d' \right\rangle  \)}  informally means ``\( P \)  with store \( d \) reduces, in one internal
step, to \( P' \)  with store \( d'\)\,''. 
We sometimes abuse of notation by writing $P \redi P'$ when $d,d'$ are unimportant. 
The \emph{observable transition} { \( P\rede{(c,d)} R \)} means ``\( P \)
on input \( c \), reduces \emph{in one time unit} to \( R \) and outputs \( d \)''. The latter is obtained from a finite sequence of internal transitions. 

In rule $R_S $, the structural congruence  \( \equiv _u  \) 
is the smallest congruence  satisfying:  
1) \( P \equiv _u Q \) if they differ only by a renaming of bound variables.  2) \( P \parallel \skipp \equiv _u P \). 3) \( P \parallel Q \equiv _u Q \parallel P \), \( P \parallel (Q \parallel R) \equiv _u (P \parallel Q) \parallel
R \). 4)  \( P\parallel \localp{\vec{x};c}{Q} \equiv _u \localp{\vec{x};c}{(P \parallel Q)} \) if $\vec{x} \not\in \fv(P)$. 5) $\localp{\vec{x};c}{}\localp{\vec{y};d}{P} \equiv _u 
	\localp{\vec{x};\vec{y}\  ; c\wedge d}{P}$ \ \ if $\vec{x}\cap\vec{y}=\emptyset $ and $\vec{y} \notin \fv(c)$. 
 Extend $\equiv _u$ by decreeing that $\mconf{P,c}\equiv _u \mconf{Q,c}$ iff $P\equiv _u Q$.

\begin{definition}[Output Behavior]\label{def:out-behav}
Let $s=c_1.c_2....c_n$ be a sequence of constraints. If $P=P_1\rede{(\true,c_1)}P_2 \rede{(\true,c_2)} \dots P_n \rede{(\true,c_n)} P_{n+1} \equiv_u Q$ we shall write $P\rede{s}^{*} Q$. If $s=c_1.c_2.c_3...$ is an infinite sequence, we omit  $Q$ in  $P\rede{s}^{*}Q$. 
The \emph{output behavior} of $P$ is defined as  $\obehav{P} = \{s \ | \ P\rede{s} ^{*}\}$.  If $\obehav{P} = \obehav{Q}$ we shall write $P \oequiv{} Q$. 
Furthermore, if $P\rede{s} Q$ and $s$ is unimportant we simply write $P\rede{}^{*}Q$. 
\end{definition}

\begin{figure}
\centering
\begin{tabular}{c}
\\
\scriptsize

$ \begin{array}{l}
\hline\\
 \rightinfer[\rTell]{\left\langle
\tellp{c},d\right\rangle \: \redi_{{}} \: \left\langle 
\skipp,d \wedge c\right\rangle}{ } 

\tab\tab\tab 

\rightinfer[\rPar]{\left\langle P\parallel
Q,c\right\rangle \: \redi_{{}} \: \left\langle P'\parallel
Q,d\right\rangle }{\left\langle P,c\right\rangle \redi_{{}}
\left\langle P',d\right\rangle}

\tab\tab\tab 

\rightinfer[\rUnless]{\left\langle \mathbf{unless}\textrm{
}c\textrm{ }\mathbf{next}\textrm{ }P,d\right\rangle  \: \redi_{{}} \:
\left\langle \mathbf{skip},d\right\rangle }{d\entails c}
\\\\ 

\rightinfer[\rLocal]{\left\langle \localp{\vec{x};c}{P},d \right\rangle \: \redi_{{}} \:
\left\langle \localp{\vec{x};c'}{P'},d\wedge \exists {\vec{x}}c'\right\rangle }
{\left\langle P, c \wedge (\exists {\vec{x}} d) \right\rangle  \:
\redi \:
\left\langle P', c' \wedge (\exists{\vec{x}} d) \right\rangle \ \ {} }

\tab\tab\tab 

\rightinfer[\rAbs]
{\tuple{\absp{\vec{x}}{c}{P}}{d}
\redi  \tuple{ P[\vec{t}/\vec{x}] || \absp{\vec{x}}{c \wedge \vec{x} \not\doteq \vec{t}\  }{P}}{d} 
} 
{   d \entails c[\vec{t}/\vec{x}] \ \ \ \ \ \ \  [\vec{t}/\vec{x}] \mbox{ is admissible }}

\\\\ 

\ \rStruct \ \frac{\raisebox{.1cm}{$\gamma_1 \redi \gamma_2$}}
{\raisebox{-.2cm}{$\gamma_1' \redi \gamma_2'$}} \ \mbox{ if }
{\gamma_1 \equiv_u \gamma_1'} \mbox{ and } {\gamma_2\equiv_u \gamma_2'} 

\tab\tab\tab 

 \rightinfer[\rBang]{\left\langle
\bangp{P},d\right\rangle \: \redi_{{}} \: \left\langle 
P || \nextp{\bangp{P}},d \right\rangle}{ }

\\\\ 
\hline \\

\ \rObserv \ \frac{\raisebox{.1cm}{$\left\langle
P,c\right\rangle \: \redi_{{}}^{*}\, \left\langle Q,d\right\rangle \:
\not\redi$}}
{\raisebox{-.2cm}{$P\rede{(c,d)} F(Q)$}} \ \mbox{ where \ \ }

{F}(P)=\left\{ 
\begin{array}{ll} 
     \skipp & \mbox{ if $  P=\skipp $ or $P=\absp{\vec{x}}{c}{Q} $}
\\   {F}(P_{1})\parallel {F}(P_{2})  & \mbox{ if $P=P_{1}\parallel P_{2} $} 
\\   \localp{\vec{x}}{F(Q)}  & \mbox{ if $ P=\localp{\vec{x};c}{Q} $ }
\\   Q      & \mbox{ if $ P=\nextp{Q} $ or $ P=\unlessp{c}{Q} $}
\end{array} \right. 

\\\\
\hline
\end{array}$
\\
\end{tabular}
\caption{\label{opersem} {
Operational Semantics for \utcc.   In $\rAbs$,  $\vec{x} \not\doteq \vec{t}$ ($\vec{x}$ syntactically different from $\vec{t}$) 
denotes  $\bigvee_{1\leq i \leq |\vec{x}|}x_i\not\doteq t_i$. If $|\vec{x}| =0$, 
$\vec{x} \not\doteq \vec{t}$ is defined as $\false$.} 
}
\end{figure}

\paragraph{Logic Correspondence.}
Remarkably, in addition to this operational view, 
\utcc processes admit a declarative interpretation based on 
Pnueli's first-order linear-time temporal logic (FLTL) \cite{mp91}. 
This is formalized by the encoding below, which maps \utcc\ processes into FLTL formulas.

\begin{definition}\label{proc2ltl:def} Let $\Logic\os \cdot \cs$ a map from $\utcc$ processes to FLTL formulas given by:
{\footnotesize
\[
\begin{array}{lllllllllll}
  \Logic\os  \skipp \cs   \!\!&=&\!\!  \true  &   &   \Logic\os \tellp{c}  \cs  \!\!&=&\!\!  c  \\
  \Logic\os P \parallel Q \cs \!\!&=&\!\!  \Logic\os P \cs \wedge \Logic\os Q \cs & & 
   \Logic\os \absp{\vec{y}}{c}{P} \cs \!\!&=&\!\!  \forall \vec{y}(c \imply \Logic\os P
  \cs) \\
     \Logic\os \localp{\vec{x};c}{P} \cs  \!\!&=&\!\!  \exists \vec{x} (c \wedge \Logic\os P \cs)  &   & \Logic\os \nextp{P} \cs  \!\!&=&\!\!\nextt \Logic\os P \cs \\ 
  \Logic\os \unlessp{c}{P} \cs  \!\!&=&\!\! c \lor \nextt \Logic \os P \cs &  & \Logic\os \bangp{P}\cs \!\!&=&\!\! \Box \Logic\os P \cs \\  
\end{array}
\]
}
\end{definition}

Modalities $\nextt F$ and $\always F$ represent that $F$ holds \emph{next} and \emph{always}, respectively.  
We use the \emph{eventual} modality $\sometime F$ as an abbreviation of $\neg \always \neg F$.

The following theorem 
relates the operational view of processes with their logic interpretation.

\begin{theorem}[Logic correspondence \cite{Olarte:08:SAC}] \label{theo:FLTL}
Let $\Logic\os \cdot \cs$ be as in Definition \ref{proc2ltl:def}, $P$ a \utcc\ process  and  $s=c_1.c_2.c_3...$ an infinite  sequence of constraints s.t.  $P\rede{s}^{*}$. For every constraint  $d$, it holds that: 
$ \  \Logic\os  P \cs  \entails  \sometime d \ \mbox{ iff  there exists $i\geq 1$ s.t. $c_i \entails d$}\,$. 
\end{theorem}

Recall that an observable transition $P\rede{(c,c')}Q$ is obtained from a finite sequence of internal transitions (rule $\rObserv$). 
We notice that there exist processes that may produce infinitely many internal transitions and as such, they cannot exhibit an observable transition; an example is $\absp{x}{c(x)}{\tellp{c(x+1)}}$.  The \utcc\ processes considered in this paper are \emph{well-terminated}, i.e., they never produce an infinite number of internal transitions during a time unit. Notice also that in the Theorem  \ref{theo:FLTL} the  process $P$ is assumed to be able to output a constraint $c_i$ for all time-unit $i\geq 1$. Therefore, $P$ must be a well-terminated process.
 
\paragraph{Derived Constructs.}
Let $\outp$ be an uninterpreted predicate.
One could attempt at representing the actions of sending and receiving as in a name-passing calculus (say, $k\bangp[\vec{e}]$ and $\receive{k}{\vec{x}}P$, resp.) with 
the \utcc processes $\tellp{\outp(k,\vec{e})}$ and $\absp{\vec{x}}{\outp(k,\vec{x})}{P}$, respectively. 
Nevertheless, since these  processes  are not automatically transferred from one time unit to the next one, they will disappear right after the current time unit, even if they do not interact.  
To 
cope with this kind of behavior, 
we shall define versions of 
$\absp{\vec{x}}{c}{P}$ and $\tellp{c}$ processes that are \emph{persistent in time}.
More precisely, 
we shall use the process $\waitpp{\vec{x}}{c}{P}$, 
which transfers itself from one time unit to the next one until, for some $\vec{t}$, 
 $c[\vec{t}/\vec{x}]$ is entailed by the current store. 
Intuitively, the process behaves like an input that is active until interacting with an output. 
When this occurs, the process outputs the constraint $\overline{c}[\vec{t}/\vec{x}]$, 
as a way of acknowledging the successful read of $c$. 
When $|\vec{x}| = 0$, we shall write $\wheneverp{c}{P}$ instead of 
$\waitpp{\vec{x}}{c}{P}$.
Similarly, we define  $\tellpp{c}$ for the persistent  output of  $c$  until some process ``reads" $c$. 
These processes can be expressed in the basic \utcc syntax as follows (in all cases, we assume $stop, go \notin \fv(c)$):

{\footnotesize
\[
\begin{array}{llll}
\tellpp{c} & \defsymbol & \localp{go,stop}{} (&
		 \tellp{\outp'(go)} \parallel \bangp\whenp{\outp'(go)}{\tellp{c}} \parallel \\
		& & &  \bangp \unlessp{\outp'(stop)}{\tellp{\outp'(go)}} \parallel  \\		
		 & & & \bangp \whenp{\overline{c}}{\bangp{\tellp{\outp'(stop)}}}) \\
		\waitp{\vec{x}}{c}{P} & \defsymbol & 
 \localp{stop,go}{}(& \tellp{\outp'(go)}  \parallel \bangp{}\unlessp{\outp'(stop)}{\tellp{\outp'(go)}} \\
 & & & \parallel \bangp   \absp{\vec{x}}{c\wedge \outp'(go)}{(P \parallel\bangp{\tellp{\outp'(stop)}}}  )	\\
 		\waitpp{\vec{x}}{c}{P} & \defsymbol & \multicolumn{2}{l}{ \waitp{\vec{x}}{c}{(P \parallel \tellp{\overline{c}})} }
		\end{array}
\] 
}

Notice that once a pair of processes $\underline{\mathbf{tell}}$ and $\underline{\mathbf{wait}}$ interact, their continuation in the next time unit is a  process able to output only  a constraint of the form 
$\exists_{x} \outp'(x)$ (e.g., $\exists_{stop}(\outp'(stop))$). 
We define the following equivalence relation that allows us to abstract from 
these processes.

\index{$\obsequiv{}$}
\begin{definition}[Observables]\label{def:obs:equivalence}
Let $\oequiv{}$ be the output equivalent relation in Definition \ref{def:out-behav}. 
We say that $P$ and $Q$ are observable equivalent, notation
$P\obsequiv{} Q$,   if $P \parallel \bangp\tellp{\exists_x{\outp'(x)}} \oequiv{}  Q \parallel \bangp\tellp{\exists_x{\outp'(x)}} $.

\end{definition}

Using the previous equivalence relation, we can show the following. 

\begin{proposition}\label{prop-wait-tell}
Assume that $c(\vec{x})$ is a predicate symbol of arity $|\vec{x}|$. 
\begin{enumerate}
	\item If $d \notentails c[\vec{t}/\vec{x}]$ for any $\vec{t}$ then $\waitpp{\vec{x}}{c}{P} \rede{(d,d)} \waitpp{\vec{x}}{c}{P} $.
	\item  If $P \equiv_u \tellpp{c(\vec{t})} \parallel \waitpp{\vec{x}}{c(\vec{x})}{\nextp Q}$ then $P \rede{} \obsequiv{}Q[\vec{t}/\vec{x}]$.
\end{enumerate}
\end{proposition}

\section{A Declarative Interpretation for Structured Communications}
\label{s:decla}

The encoding $[|\cdot|] $ from \hvk into \utcc is defined in Table \ref{table:enc}.
Two noteworthy aspects when considering such a translation are \emph{determinacy} and \emph{timed behavior}. 
Concerning determinacy, it is of uttermost importance to 
recall that 
while \utcc\ is a deterministic language, 
\hvk\ processes may exhibit non-deterministic behavior. 
Moreover, while 
\hvk is a synchronous language, whereas \utcc is asynchronous. 
Consider, for instance, the \hvk process:
\[
P =\send{k}{\vec{e}}Q_1 \ | \ \send{k}{\vec{e'}}Q_2 \ | \  \receive{k}{\vec{x}}Q_3
\]
Process $P$ can have two possible transitions,
and evolve into  $\send{k}{\vec{e'}}Q_2 \ | \ \ Q_3[\vec{e}/\vec{x}]$
or into $\send{k}{\vec{e}}Q_1 \ | \ \ Q_3[\vec{e'}/\vec{x}]$.
In both cases, there is an output that cannot interact with the input  $\receive{k}{\vec{x}}Q_3$.
In \utcc, inputs are represented by abstractions which are persistent during a time unit. 
As a result, in the encoding of $P$ we shall observe that \emph{both} outputs react with the same input, i.e. that
$\os P \cs \rede{} \os Q_3[\vec{e}/\vec{x}]\cs \parallel \os Q_3[\vec{e'}/\vec{x}]\cs$. 

As for timed behavior, 
it is crucial to observe that while 
\hvk is an untimed calculus, 
\utcc provides constructs for explicit time. 
In the 
encoding we 
shall advocate 
a timed interpretation of \hvk in which all available synchronizations between processes occur at a given 
time unit,
and the continuations of synchronized processes will be executed in the next 
time unit.
This will prove convenient when showing the operational correspondence between both calculi, as we can relate the observable behavior in \utcc and the reduction semantics in \hvk.
 
Let us briefly provide some intuitions on $[|\cdot|]$. 
Consider \hvk processes $P=\request{a}{k}P'$ and $Q=\accept{a}{x}Q'$. 
The encoding of $P$ declares a  new variable session $k$ and sends it through the channel $a$ by posting the constraint $\reqp(a,k)$. Upon reception of the session key (local variable) generated by $\os P\cs$, process $\os Q\cs$ adds the constraint $\accp(a,k)$
to notify the acceptance of $k$. 
They 
can then synchronize on this constraint, and execute their continuations in the next time unit. 
 The encoding of label selection and branching  
is similar, and uses constraint $\selp(k,l)$ for synchronization. 
We use the parallel composition 
$\prod\limits_{1\leq i \leq n} \whenp{l=l_i}{\nextp{\os P_i\cs}}$ to execute the selected choice. Notice that we do not require a non-deterministic choice since the constraints $l=l_i$ are mutually exclusive.  
As in \cite{honda1998lpa}, in the encoding of  $\itn{e}{P}{Q}$ we assume an evaluation function on expressions. Once $e$ is evaluated, $\downarrow e$ is a \emph{constant} boolean value. 
The encoding of  $\recursionH{D}P$ 
exploits the scheme described in Equation \ref{eq:def-encoding}. 

\begin{table}[t]
\centering
{\footnotesize
\begin{longtable}{c}
\\ 
$
\begin{array}{r c l}
[| \request{a}{k}P  |] & =	&	 
	\localp{k}{}( \tellpp{\reqp(a,k)} \parallel \wheneverp{\accp(a,k)}{\nextp} \os P \cs) \\ 
	  
	  [|\accept{a}{k}P |] &=  & \waitpp{k}{\reqp(a,k)}{}(\tellp{\accp(a,k)} \parallel  \nextp \os P \cs) \\ \\
%
	  
	  [| \send{k}{\vec{e}}P |] & = & \tellpp{\outp(k,\vec{e})} \parallel \wheneverp{\overline{\outp(k,\vec{e})}}\nextp{\os P \cs}  \\  
	  
	  [| \receive{k}{\vec{x}}P |] &=& \waitpp{\vec{x}}{\outp(k,\vec{x})}{\nextp{\os P \cs}} \\ \\
	  
	  [| \selection{k}{l}P |] &= & \tellpp{\selp(k,l)} \parallel \wheneverp{\overline{\selp(k,l)}}{\nextp{\os P \cs}} \\ 
	  
	  [| \branching{k}{l_1:P_1 || \dots || l_n:P_n} |] & = & \waitpp{l}{\selp(k,l)}{
	  \prod\limits_{1\leq i \leq n} \whenp{l=l_i}{\nextp{\os P_i\cs}}
	  } \\ 
	  
	  [| \throw{k}{k'}P |] & = & \tellpp{\outkp(k,k')} \parallel \wheneverp{\overline{\outkp(k,k')}}{\nextp{\os P \cs}} \\ 
	  
	  [| \catch{k}{k'}P |] & = & \wheneverpp{\outkp(k,k')}{\nextp{\os P \cs}} \\ \\
	  
	  [| \itn{e}{P}{Q} |] & = & \whenp{e\downarrow\true}{\nextp{\os P \cs}} \parallel \whenp{e\downarrow\false}{\nextp{\os Q \cs}} \\ 
	  
	   [| P | Q |]	   & = & \os P \cs \parallel \os Q \cs \\
	   
	   [| \inact|] & =	&	\sskip \\ 
	  
	   [| (\nu u) P|] & =	&	\localp{u}{ \os P \cs} \\ 	
	  \os \recursionH{D}P \cs	&	=&	\prod\limits_{X_i(x_ik_i) \in D } \cR\os X_i(x_ik_i)\cs {\widehat{P}} \\
	   \end{array}
$
\end{longtable}
}
\caption{{An Encoding from \hvk into \utcc. $\cR \os \cdot \cs$ and $\widehat{P}$ are defined in Equation \ref{eq:def-encoding}. \label{table:enc}}}
\end{table} 


\paragraph{Operational Correspondence.}
Here we study an operational correspondence property for our encoding. 
The differences with respect to (a)synchrony and determinacy discussed above 
will have a direct influence on the correspondence. 
Intuitively, the encoding falls short for \hvk programs featuring the kind of non-determinism that results from ``uneven pairings'' between session requesters/providers, label selection/branching, and inputs/outputs as in the example above. 

We thus find it convenient to  appeal to the type system of \hvk\ to obtain some basic determinacy  
of the source terms.  
Roughly speaking, the type discipline in \cite{honda1998lpa} ensures a correct pairing
between actions and co-actions once a session is established. 
Although the type system guarantees a correct match between (the types of) session requesters and providers, 
it does not rule out the
kind of non-determinism induced by different orders in the pairing of requesters and providers. 
We shall then require session providers to be always willing to engage into a session. 
This is, given a channel $a$, we require that there is at most one  $\mathbf{accept}$ process (possibly replicated) on $a$  
that is able to synchronize with every process requesting a session on $a$. 
Notice that this requirement is in line with a meaningful class of programs, namely those 
described by the type discipline developed in \cite{Berger2008Completeness-an,BergerHY01}.

Before presenting the operational correspondence, we introduce some 
auxiliary notions.



\begin{definition}[Processes in normal form]\label{def:normal-form:hvk}
We say that a \hvk\ process   $P$ is in \emph{normal form} if takes the form $\inact$ or $\recursionH{D}\nu \vec{u}(Q_1 \ |\   \cdots \ |\  Q_n)$ where neither
the operators ``$\nu$'' and ``$|$''  nor process variables occur in  the top level of   $Q_1,\ldots,Q_n$.
\end{definition}

The following proposition states that given a process $P$ we can find a process $P'$ in normal form,
such that: either $P'$ is structurally congruent to $P$, or it results from 
replacing the process variables at the top level of $P$ with their corresponding definition (using rule $\textsc{Def}$). 

\begin{proposition}
For all \hvk\ process $P$ there exists $P'$ in normal form s.t. $P \redihvk^{*}\equiv _h P'$ only using the rules $\textsc{Def}$ and $\textsc{Str}$ in Figure \ref{tab:sos-hvk}.
\end{proposition}
\begin{proof}
Let $P$ be a process of the form $\recursionH{D}Q$  where there are no  procedure definitions in $Q$. By repeated applications of the rule $\textsc{Def}$, we can show that $P\redihvk^{*} P'$ where $P'$ does not have occurrences of processes variables in the top level. Then, we use the rules of the structural congruence to move the local variables to the outermost position and find $P''\equiv _h P'$ in the desired normal form.
\end{proof}

Notice that the rules of the operational semantics of \hvk\  are given for pairs of processes that can interact with each other. 
We shall refer to each of those pairs as a \emph{redex}. 

\begin{definition}[Redex]\label{d:redex}
A \emph{redex} is a pair of complementary processes composed in parallel as in:
{\footnotesize \begin{align*}
(1) &\request{a}{k}P \  |\   \accept{a}{k}Q & (3) & \throw{k}{k'}P \  | \ \catch{k}{k'}Q.\\
(2) &\send{k}{\vec{e}}P \ |\  \receive{k}{\vec{x}}Q & (4) & \selection{k}{l}P \ |\  \branching{k}{l_1:P_1 \parallel \cdots || l_n:P_n}
\end{align*} }
\end{definition}



Notice that  a redex in \hvk\  synchronizes and reduces in a single transition as in $ (\send{k}{\vec e}P) \ |\  (\receive{k}{\vec{x}}Q )$ $\redihvk P \ |\  Q [\vec{e}/\vec{x}] $.
Nevertheless, in \utcc, the encoding of the processes above requires several internal transitions  for adding the constraint $\outp(k,\vec{e})$ to the current store, and for ``reading" that constraint by means of  $\waitpp{\vec{x}}{\outp(k,\vec{x})}{\nextp \os Q \cs}$ to later execute $\nextp \os Q[\vec{e}/\vec{x}]\cs$.
We shall then establish the  operational correspondence between an observable transition of \utcc (obtained from a finite number of internal transitions) and the following subset of reduction relations over \hvk\ processes:

\begin{definition}[Outermost Reductions]
Let  $P\equiv_h \recursionH{D}\nu \vec{x}(Q_1 \,|\,\cdots \, | \,Q_n)$ be an \hvk\ program in normal form. 
We define the \emph{outermost reduction relation} $P\rede{}_{h}P'$ as the maximal sequence of  reductions
$ P \redihvk^{*} P' \equiv _h \recursionH{D}\nu\vec{x'}(Q_1' \, |\, \cdots \, | \, Q_n')$ such that for every $i\in\{1,..n\}$, either 
\begin{enumerate}
\item $Q_i = \itn{e}{R_1}{R_2} \redihvk R_{1/2} = Q_i'$; 
\item for some $j\in\{1,..n\}$, $Q_i | Q_j$ is a redex 
such that $Q_i | Q_j \redihvk \nu\vec{y}(Q_i' | Q_j')$, with $\vec{y} \subseteq \vec{x'}$; 
\item there is no $k\in\{1,..n\}$ such that $Q_i \, | \, Q_k$ is a redex and $Q_i \equiv _h Q_i'$.

\end{enumerate}
\end{definition}

One may argue that the above-presented definition may rule out some possible reductions in \hvk. Returning to the concerns about determinacy, an outermost reduction filters out cases where there are more than one possible reduction for a set of parallel processes (i.e.: the parallel composition of two outputs and one input with the same session key). The use of outermost reductions gives us a subset of possible reductions in \hvk that keeps synchronous processes and discard processes that are not going to interact in any way (recall that in the typing discipline of  \hvk the composition of an input and an output with the same session key will consume the channel used; hence, every other process sending information over the same session will not have any complementary process to synchronize with).

In the sequel we shall thus consider only  \hvk\ processes $P$ where for  $n\geq 1$, if $P\equiv_h P_1\rede{}_{h}P_2 \rede{}_{h} \cdots \rede{}_{h} P_n$ and $P\equiv_h P_1'\rede{}_{h}P_2' \rede{}_{h}\cdots \rede{}_{h} P_n'$ then $P_i \equiv _h P_i'$ for all $i\in \{1,..,n\}$, i.e., $P$ is a \emph{deterministic} process.

\begin{theorem}[Operational Correspondence] \label{thm:oper-corr}
Let $P,Q$ be deterministic \hvk\  processes in normal form and $R$,$S$ be \utcc\ processes. It holds: \\
1) {\em Soundness}:   If  $P \rede{}_{h}  Q$ then, for some $R$,  $\os P \cs \rede{} R \obsequiv{} \os Q \cs$; \\
 2) {\em Completeness}:  If $ \os P\cs  \rede{} S$ then, for some $Q$, $P \rede{}_{h} Q$ and $\os Q\cs  \obsequiv{} S$.
\end{theorem}
\begin{proof}
Assume that $P\equiv _h \recursionH{D}\nu \vec{x}(Q_1 \,|\,\cdots \, | \,Q_n)$ 
and $Q\equiv _h \recursionH{D}\nu \vec{x'}(Q_1' \,|\,\cdots \, | \,Q_n')$.

\begin{enumerate}
	\item {\em Soundness}. 
	Since $P \rede{}_{h} Q$ there must exist a sequence of derivations of the form 
	$P \equiv _h P_1 \redihvk P_2 \redihvk ... \redihvk P_n \equiv _h Q$. 	The proof
proceeds by induction on the length of this derivation, with a case analysis on the last applied rule. We then have the following cases:
	
	\begin{enumerate}
		\item {\bf Using the rule \textsc{If1}}. It must be the case  that 
		there exists $Q_i \equiv _h \itn{e}{R_1}{R_2}$ and $Q_i \redihvk R_{1} \equiv _h Q_i'$ and $e\downarrow \true$. One can easily show that  $\whenp{e \downarrow \true}\nextp$ $\os Q_i'\cs \rede{}\os Q_i'\cs$. 

		\item {\bf Using the rule \textsc{If2}} Similarly as for \textsc{If1}.

		\item {\bf Using the rule \textsc{Link}}. It must be the case that 
		there exist $i,j$ such that $Q_i \equiv _h \request{a}{k}Q_i'$ and 
		$Q_j \equiv _h \accept{a}{x}Q_j'$ and then 
		$Q_i \ | \ Q_j \redihvk (\nu k)(Q_i' \ | \ Q_j')$. We then have a derivation
		\[\footnotesize
		\begin{array}{lll}
		\os Q_i\cs \parallel \os Q_ k\cs & \redi^{*} &
		\localp{k;c}{}( R_i' \parallel \wheneverp{\accp(a,k)}{\nextp} \os Q_i' \cs \parallel \\
		& & \tab\tab\tab\tab\ \ \waitpp{k'}{\reqp(a,k')}{}(\tellp{\accp(a,k')} \parallel  \nextp(\os Q_j' \cs)) \\
		& \redi^{*} & 		
		\localp{k;c'}{}( R_i' \parallel \wheneverp{\accp(a,k)}{\nextp} \os Q_i' \cs \parallel \\
		& & \tab\tab\tab\tab\ R_j' \parallel \tellp{\accp(a,k)} \parallel  \nextp(\os Q_j'[k/k'] \cs) \\
		& \redi^{*} & 		
		\localp{k;c''}{}( R_i' \parallel R_j' \parallel \nextp \os Q_i' \cs \parallel  \nextp(\os Q_j'[k/k'] \cs) \not\redi\\

		\end{array}
		\]
		
		where 
		${\small
		c  =  \reqp(a,k),  \ 
	 	c'  = c  \wedge  \overline{\reqp(a,k)}, \ 
		c''  =  c'  \wedge \accp(a,k) \wedge 
		\overline{\accp(a,k)} 
		}
		$
	and 
	$R_i'$, $ R_j'$ are the processes resulting after the interaction of the processes in the parallel composition
	$\tellpp{\reqp(a,k)} \parallel \waitpp{k'}{\reqp(a,k')}{\cdots}$,  i.e.:\[\footnotesize
\begin{array}{rll}
R_i'& \equiv_u & \localp{go,stop ; \outp'(go)\wedge \outp'(stop) \wedge c(\vec{t})}{}\\
		& & \tab\tab\tab\tab\tab \nextp\bangp \unlessp{\outp'(stop)}{\tellp{\outp'(go)}} \parallel \nextp {\bangp{\tellp{\outp'(stop)}}}\\
R_j' & \equiv_u & \localp{stop',go' ; \outp'(go') \wedge \overline{c}(\vec{t}) \wedge \outp'(stop')  }{} \nextp\bangp\tellp{\outp'(stop')}  \\ 
& & \ \  \parallel  \nextp \bangp{}\unlessp{\outp'(stop')}{\tellp{\outp'(go')}} \\
& &  \ \ \parallel   \absp{\vec{x}}{c\wedge \outp'(go') \wedge \vec{x} \not\doteq \vec{t}}{(Q \parallel \tellp{\overline{c}(\vec{t})} \parallel \bangp{\tellp{\outp'(stop')}}} \\
& &  \ \ \parallel \nextp\bangp   \absp{\vec{x}}{c\wedge \outp'(go') }{(Q \parallel \tellp{\overline{c}(\vec{t})} \parallel \bangp{\tellp{\outp'(stop')}}}  
\end{array}\]

We notice that  $R_i' \parallel R_j' \notredi$ and it is a process that  	can only output the constraint $\outp'(x)$ where $x$ is a local variable. By appealing to Proposition \ref{prop-wait-tell} we conclude 
		$\os Q_i\cs \parallel \os Q_ j\cs \rede{}\obsequiv{} \localp{k}{}(\os Q_i'\cs \parallel \os Q_ j'\cs$).
		
		\item The cases using the rules $\textsc{Label}$ and $\textsc{Pass}$ can be proven similarly as the case for 
		$\textsc{link}$.
	\end{enumerate}
	
	\item {\em Completeness}. Given the encoding and the structure of $P$, we have a \utcc process $R = \os P \cs$ s.t.
	\[
	R \equiv _u \localp{\vec{x}}(\os Q_1 \cs \parallel ... \parallel \os Q_n \cs) \, .
	\]
	
	Let $R_i = \os Q_i \cs$ for $1\leq i \leq n$. 
 By an analysis on the structure of $R$, if $R_i \redi R_i'$ then it must be the case that either (a)
 $R_i= \whenp{e}\nextp{\os Q_i'\cs}$ and  $R_i' = \nextp \os Q_i'\cs$ or (b) $\langle R_i , c\rangle \redi \langle R_i',c \wedge d \rangle$
  where $d$ is a constraint of the form  $\reqp(\cdot)$, $\selp(\cdot)$, $\outp(\cdot)$, or $\outkp(\cdot)$.
	In both cases we shall show that there exists a $R_i''$ such that $R_i \redi^{*} R_i'' \notredi$  such that 
	$Q_i \redihvk Q_i' $ and  $R_i'' = \nextp  \os Q_i' \cs$. 
	
	
	\begin{enumerate}
		
			\item Assume that $R_i = \whenp{e\downarrow \true} \nextp \os Q_i' \cs$ for some $Q_i'$.  
			Then it must be the case that 
			$Q_i=\itn{e}{Q_i'}{Q_i''}$.
			If $e\downarrow\true$  we then have 
			$R_i'' = \nextp \os Q_i'\cs$. 
			The case when $e \downarrow \false$ is similar by considering $R_i = \whenp{e\downarrow \false}{Q_i'}$.

		\item Assume now that $\langle R_i,c \rangle \redi \langle R_i' , c \wedge d \rangle$ where 
		$d$  is of the form $\reqp(\cdot)$, $\selp(\cdot)$, $\outp(\cdot)$ or $\outkp(\cdot)$. 
			We proceed by case analysis of the constraint 
		$d$. Let us consider only the case $d = \exists_k(\reqp(a,k))$; the cases in which $d$ takes the form $\selp(\cdot)$, $\outp(\cdot)$, or $\outkp(\cdot)$ are handled similarly. 
If $d = \exists_k(\reqp(a,k))$ for some $a$, then we must have that  $Q_i \equiv _h \request{a}{k}Q_i'$ for some $i$. 
	If there exists $j$ such that 
			$Q_j \equiv _h \accept{a}{x}Q_j'$, one can show a derivation similar to the case of the rule $\textsc{Link}$ in soundness to prove that $R_i \parallel R_j \redi^{*} \oequiv{} \localp{k}(\nextp \os Q_i'  \cs \parallel  \nextp \os  Q_j' \cs)$.
		If there is no $Q_j$ such that $Q_i \ | Q_j$ forms a redex,
	then one can show by using (1) in Proposition \ref{prop-wait-tell} that $R_i \rede{} \obsequiv{} R_i$ .
	\end{enumerate}	
\end{enumerate}
\end{proof}

\section{A Timed Extension of \hvk}

We now propose an extension to \hvk in which 
a bundled treatment of time is explicit and session closure is considered. 
More precisely, the \hvkplus language arises as the extension of \hvk processes (Def. \ref{def-hondasLanguage}) 
with refined constructs for session request and acceptance, as well as with a construct for
session abortion: 

\begin{definition}[A timed language for sessions]
\hvkplus processes are given by the following syntax:

\begin{center}
{\footnotesize
\begin{tabular}{rrll} 
	P	&$::=$	& 	\timedRequest{a}{k}{m}P  	&	Timed Session Request\\
			&$|$	&	\timedAccept{a}{k}{c}P	     & Declarative Session Acceptance\\
                        &$|$    &       $\cdots$  & \{~~the other constructs, as in Def. \ref{def-hondasLanguage} ~~\} \\
			&$|$	&	\killS{c_k}	&	Session Abortion\\
 \end{tabular}
}
\end{center}
\end{definition}

The intuition behind these three operators is the following: $\timedRequest{a}{k}{m}P$ will request a session $k$ over the service name $a$ during $m$ time units. Its dual construct is $\timedAccept{a}{k}{c}P$: 
it will grant the session key $k$ when requested over the service name $a$ 
provided by a session and a successful check over the constraint $c$. 
Notice that $c$ stands for a precondition for agreement between session request and acceptance. In $c$, 
the duration  $m$ of the corresponding session key $k$
can be referenced by means of the variable ${\it dur_k}$. In the encoding we syntactically replace it by the  variable corresponding to $m$. 
Finally, \killS{c_k} will remove $c_k$ from the valid set of sessions.

\begin{table}[h]
\centering
{\footnotesize
\begin{tabular}{c}
\\
$
\begin{array}{r c l}
[| \timedRequest{a}{k}{m}P  |] & =	&	 
	\localp{k}{}\tellpp{\reqp(a,k,m)} \parallel  \\
	& & \ \ \tab\tab\tab  \wheneverp{\accp(a,k)}{\nextp}( \tellp{\actp(k)} \parallel \Guarded_{\actp(k)} (\os P \cs ) \parallel \\
	  & & \qquad \qquad \qquad \qquad \qquad \qquad \qquad \tab \ \ \bangp_{[m]}\unlessp{\killp(k)} \tellp{\actp(k)}) \\ 
	  
	  [|\timedAccept{a}{k}{c}P |] &=  & \waitpp{k}{\reqp(a,k,m) \land c[m/{\it dur_k}]}{}(\tellp{\accp(a,k)} \parallel  \nextp\Guarded_{\actp(k)}(\os P \cs)) \\ 
\os \killS{k} \cs & =	&	\bangp\tellp{\killp(k)} \\ \\
	   \end{array}
$\\
\end{tabular}
}
 \caption{{\label{table:enc2} The Extended Encoding. $\Guarded_d(P)$ is in Definition \ref{def:guarded}.}}
 \label{enc:ext}
\end{table} 

Adapting the encoding in Table \ref{table:enc}  to consider \hvkplus processes 
is remarkably simple (see Table \ref{enc:ext}). 
Indeed, modifications to the encoding of session request and acceptance are straightforward.
The most evident change is the addition of the parameter $m$ within the constraint $\reqp(a,k,m)$.
The duration of the requested session is suitably represented as a bounded replication of the process defining the activation of
the session $k$ represented as the constraint $\actp(k)$. The execution of the continuation $\os P \cs$ is guarded by the constraint $\actp(k)$ (i.e. $P$ can be executed only when the session $k$ is valid). 
Thus, in the encoding we use the function $\Guarded_d(P)$ to denote the process behaving as $P$ when the constraint $d$ can be entailed from the current store, doing nothing otherwise. More precisely:

\begin{definition}\label{def:guarded}
 Let  $\Guarded: \cC \to Procs \to Procs$ be defined as 
{ \footnotesize
\begin{align*}
&\Guarded_d(\skipp) &=&  \skipp  & & \Guarded_d( P_1 \parallel P_2)  &=& \Guarded_d(P_1) \parallel \Guarded_d(P_2)  \\
&\Guarded_d(\tellp{c}) &=& \whenp{d}{\tellp{c}}   & & \Guarded_d( \bangp{Q}) &=& \bangp\Guarded_d(Q) \\
&\Guarded_d( \nextp{Q}) &=& \whenp{d}{\nextp\Guarded_d(Q)}   & &  \Guarded_d(\absp{\vec{x}}{c}{Q}) &=& {\absp{\vec{x}}{c}{\Guarded_d(Q)}}  \quad \text{ if $\vec{x} \notin \fv(d)$}  \\
&\Guarded_d( \unlessp{c}{Q}) &=& \whenp{d}{\unlessp{c}{\Guarded_d(Q)}}  & &  \Guarded_d( \localp{\vec{x};c}{Q}) &=& \localp{\vec{x};c}{\Guarded_d(Q)}   \quad \text{ if $\vec{x} \notin \fv(d)$}
\end{align*}}
\end{definition}

On the side of session acceptance, the main novelty is the introduction of $c[m/{\it dur_k}]$. As explained before, we syntactically replace the variable ${\it dur_k}$ by the corresponding duration of the session $m$. 
This is a generic way to represent the agreement that should exist between a service provider and a client; for instance, it could be the case that the 
client is requesting a session longer than what the service provider can or want to grant.

\subsection{Case Study: Electronic booking}
Here we present 
an example that 
makes use of the constructs introduced in \hvkplus. 

Let us consider an electronic booking scenario. 
On one side, consider 
a company AC which offers flights directly from its website. On the other side, there is a customer looking for the best offers. 
In this scenario, the customer establishes a timed session with AC and asks for a flight proposal given a set of constraints (dates allowed, destination, etc.). After receiving an offer from AC, the customer can refine  the selection further (e.g. by checking that the prices are below a given threshold) and loops until finding a suitable option, that he will accept by starting the booking phase. 
One possible \hvkplus specification of this scenario is described in Table \ref{t:ex1}.

\begin{table}[h]
\footnotesize{
\centering
\begin{tabular}{lcl} 
\\
Customer & =  & $\timedRequest{ob}{k}{m} (\send{k}{booking data} Select(k))$\\ 
Select(k)	& = &~ $\receive{k}{offer} ( \itn{(offer.price \leq 1500) }{\selection{k}{Contract}}{Select(k)})$ \\ 
AC	& = &	$\timedAccept{ob}{k}{dur_k \leq MAX\_TIME} ( \receive{k}{bookingData} $ \\
	&	&	~~ $(\nu u) \send{k}{u} \branching{k}{Contract: \overline{Accept} || Reject : \killS{k}})$\\
\end{tabular} \caption{{Online booking example with two agents. \label{t:ex1}}}
}
\end{table}
%

In a second stage, the customer uses an online broker to mediate between him and a set of airlines acting as service providers. Let $n$ be the number of service providers, and consider two vectors of fixed length: ${\it Offers}$, which contains the list ${\it [Offers_0,\dots, Offers_i,\dots, Offers_n]}$ of offers received by a customer, and $SP$, which contains the list of trusted services. First, the customer establishes a session with the broker for a given period $m$; later on, he/she starts requesting for a flight by providing the details of his/her trip to the broker. On the other side, 
the broker 
will look into his pool of trusted service providers for the ones 
that can supply 
flights 
that suit the customer's requirements.
All possible offers are transferred back to the customer, who will invoke a local procedure $Sel$ (not specified here) that selects one of the offers by performing an output on name $a$. 
Once an offer is selected, the broker will allow a final interaction between the  customer and the selected service. He does so by  delegating to the customer the session key used previously between him and the chosen service provider. 
Finally, the broker  proceeds to cancel all those sessions concerning the discarded services.
An \hvkplus specification of this scenario is given in 
Table  \ref{t:ex2} 
where,  for the sake of readability, processes denoting post-processing activities are abstracted from the specification.

A notable advantage in using \hvkplus as a modeling language is 
the possibility of exploiting timed constructs in the specification of 
service enactment and service cancellation. 
In the above scenario it is possible to see how \hvkplus allows (i) to effectively take explicit account on the maximal times accepted by the customer: the composition of nested services can take different speeds but the service broker will ensure that customers with low speeds are ruled out of the communication;  and (ii) to have a more efficient use of the available resources: since there is not need to maintain interactions with discarded services, the service broker will free those resources by sending kill signals.



\begin{table}[h]
\footnotesize{
\hspace{-0.3cm}
	\subfigure[Customer and Service Provider]{
\begin{tabular}{ll} 
Customer =   &\hspace{-0.4cm} $\timedRequest{ob}{k}{m} (\send{k}{booking data} $\\ 
	&\hspace{-0.5cm}  ~$\receive{k}{ n } ($\\
	&\hspace{-0.5cm}  ~$\prod\limits_{i \in n} (\receive{k}{\it Offers_i} ( $\\
	&\hspace{-0.5cm}	~$Sel({\it Offers});\receive{a}{x}\send{k}{\it x}$ \\
	&\hspace{-0.5cm} ~~$\catch{k}{k'}$\\
	&\hspace{-0.5cm} ~~$\send{k'}{PaymentDetails} \inact))))$\\\\
SP = &\hspace{-0.4cm}	$\timedAccept{SP_i}{k_i'}{N \leq 300ms} ($\\
	&\hspace{-0.5cm}	~$\receive{k_i'}{bookingData} $\\
	&\hspace{-0.5cm}	~~$\send{k_i'}{\it offer}$\\
	&\hspace{-0.5cm}	~~$\receive{k_i'}{paymentDetails} \inact)$\\
\end{tabular}
	}
	\subfigure[Online Broker]{ \hspace{-0.3cm}
\begin{tabular}{ll} 
Broker = &	\hspace{-0.4cm} $\timedAccept{ob}{k}{m \leq 500ms} ( $ \\
	&\hspace{-0.5cm}	~ $\receive{k}{bookingData} \send{k}{ |SP| }$ \\
	&\hspace{-0.5cm}	~ $(\nu u) \prod\limits_{ i \in |SP|} (\timedRequest{SP_i}{k_i'}{N} $\\
	&\hspace{-0.5cm}	\quad$\send{k_i'}{bookingData}$\\
	&\hspace{-0.5cm}	\quad $\receive{k_i'}{\it offer_i} (\send{u}{\it offer_i} \inact || S(u,k)))$\\
	&\hspace{-0.5cm}	\quad $\receive{k}{y} \recursionH{ X({\it Offers}, k_1', \dots, k_n') = P }$\\
	&\hspace{-0.5cm}	\quad \quad $ \prod\limits_{i \in |SP|} (\itn{(\it y = offers_i)}{(\throw{k}{k_i'} PostProc )}{}$\\
	 &\hspace{-0.5cm}	\quad \quad	\quad \quad ${\killS{k_i'}||P(X-\{{\it offers_i}, k_i'\})}))$\\\\
S(u,k) = &\hspace{-0.5cm} 	$\prod\limits_{ i \in |SP|} (\receive{u}{\it offer_i} \inact || \send{k}{\it offer_i} \inact )$\\
\end{tabular}
	}
 \caption{Online booking example with online broker.\label{t:ex2}}
}
\end{table}
\subsection{Exploiting the Logic Correspondence}
To exploit the logic correspondence we can draw inspiration from the \emph{constraint templates} put forward in 
\cite{pesic2006daf}, 
a set of LTL formulas that represent desirable/undesirable situations in service management. Such templates are divided in three types: \emph{existence constraints}, that specify the number of executions of an activity; \emph{relation constraints},  that define the relation between two activities to be present in the system; and \emph{negation constraints}, which are essentially the negated versions of relation constraints. 

By appealing to Theorem \ref{theo:FLTL}, our framework allows for the verification of  existence and relation constraints over \hvkplus programs. Assume a \hvkplus program $P$ and let $F = \Logic\os  \os P \cs \cs$ (i.e., the FLTL formula associated to the \utcc\ representation of $P$). For existence constraints, assume that $P$ defines a service accepting requests on channel $a$. If the service is eventually active, then it must be the case that $F \entails \sometime \exists_k(\accp(a,k))$ (recall that the encoding of $\mathbf{accept}$ adds the constraint $\accp(a,k)$ when the session $k$ is accepted). A slight modification to the encoding of $\mathbf{accept}$ would allow us to take into account  the number of accepted sessions and then support the verification of properties such as $F \entails \sometime (N_{sessions}(a) = N)$, informally meaning that the service $a$ has accepted $N$ sessions. This kind of formulas correspond to the existence constraints  in \cite[Figure 3.1.a--3.1.c]{pesic2006daf}. Furthermore, making use of the guards associated to ask statements, we can verify relation constraints as eventual consequences over the system. Take for instance the specification in Table \ref{t:ex1}. Let  $\overline{Accept}$ be a process that outputs ``${\it ok}$" through a session $h$. We then may verify the formula $F \entails \exists_u (u.price < 1.500 =>  \outp(h, {\it ok}))$. This is a responded existence constraint describing how the presence of an offer with price less or equal than $1.500$  would lead to an acceptance state.

\section{Concluding Remarks}
\label{s:concl}
We have argued for a timed CCP language as a suitable foundation for analyzing structured communications.
We have presented 
an encoding of the language for structured communication in \cite{honda1998lpa} into \utcc, as well as an extension of such a language that considers explicitly elements of partial information and session duration.
To the best of our knowledge, a unified framework where behavioral and declarative techniques converge for the analysis of structured communications has not been proposed before.

Languages for structured communication and CCP process calculi are conceptually very different.
We have dealt with some of these differences (notably, determinacy) 
when stating an operational correspondence property for the declarative interpretation of \hvk processes. 
We believe there are at least two ways of achieving more 
satisfactory 
notions of operational correspondence.
The first one involves considering 
extensions of \utcc 
with (forms of) non-determinism. 
This would allow to capture some scenarios of session establishment in which the operational correspondence presented here falls short.
The main consequence of adding non-determinism to \utcc is that 
the correspondence with FLTL as stated in Theorem \ref{theo:FLTL} would not longer hold. This is mainly because non-deterministic choices cannot be faithfully represented as logical disjunctions (see,  e.g., \cite{NPV02}).
While a non-deterministic extension to \tcc with a tight connection with temporal logic 
has been developed 
(\ntcc \cite{NPV02}), 
it does not provide for representations of mobile links. 
Exploring whether there exists a CCP language between \ntcc and \utcc combining both non-determinism and mobility 
while providing logic-based reasoning techniques is interesting on its own and appears challenging.
The second approach consists in defining a type system for \hvk and \hvkplus processes better suited to the nature of \utcc processes.
This would imply enriching the original type system in \cite{honda1998lpa} with 
e.g., stronger typing rules for dealing with session establishment. 
The definition of such a type system is delicate and needs care, as one would not like to rule out 
too many processes as a result of too stringent typing rules. 
An advantage of a type system ``tuned" in this way is that one could aim at obtaining a correspondence between 
well-typed processes and logic formulas, similarly as the given by Theorem \ref{theo:FLTL}.
In these lines, plans for future work include the investigation of effective mechanisms for 
the seamless integration of new type disciplines and reasoning techniques based on temporal logic within the elegant framework 
provided by 
(timed) CCP languages.

The timed extension to \hvk presented here includes notions of time that involve only session engagement processes.
A further extension could involve the inclusion of time constraints over input/output actions.
Such an extension might be useful to realistically specify scenarios in which factors such as, e.g, 
network traffic and long-lived transactions, 
prevent interactions between services from occurring instantaneously.
Properties of interest in this case could include, for instance, 
the guarantee that a given interaction has been fired at a valid time, or 
that the nested composition of services does not violate a certain time frame. 
We plan to explore case studies of structured communications involving this kind of timed behavior, and 
extend/adjust \hvkplus accordingly.

\paragraph{Acknowledgments.}
We are grateful to Marco Carbone and Thomas Hildebrandt for insightful discussions on the topics of this paper.
We also grateful to Roberto Zunino who provided useful remarks on a previous version of this document.
The contribution of Olarte and P\'{e}rez was initiated during short research visits to the IT University of Copenhagen.
They are most grateful to the IT University and to the FIRST PhD Graduate School for funding such visits.


\bibliographystyle{abbrv}
\bibliography{reportes}

\appendix 

\end{document}